\documentclass[12pt,preprint]{aastex}

\shorttitle{Compact Radio Sources in NGC 253}
\shortauthors{Lenc Tingay}

\begin{document}
%
%
\title{The Sub-parsec Scale Radio Properties of Southern Starburst Galaxies.
I. Supernova Remnants, the Supernova Rate, and the Ionised Medium in the NGC 253 Starburst}
%


\author{E. Lenc \& S.J. Tingay}

\affil{Centre for Astrophysics and Supercomputing, Swinburne University of Technology,
Mail number H39, P.O. Box 218, Hawthorn, Victoria 3122; elenc@astro.swin.edu.au}
%
\begin{abstract}

Wide-field, very long baseline interferometry (VLBI) observations of the nearby starburst galaxy NGC 253, obtained with the Australian Long Baseline Array (LBA), have produced a 2.3 GHz image with a maximum angular resolution of 15 mas (0.3 pc). Six sources were detected, all corresponding to sources identified in higher frequency ($>5$ GHz) VLA images. One of the sources, supernova remnant 5.48$-$43.3, is resolved into a shell-like structure approximately 90 mas (1.7 pc) in diameter. From these data and data from the literature, the spectra of 20 compact radio sources in NGC 253 were modelled and found to be consistent with free-free absorbed power laws. Broadly, the free-free opacity is highest toward the nucleus but varies significantly throughout the nuclear region ($\tau_0\sim 1->20$), implying that the overall structure of the ionised medium is clumpy. Of the 20 sources, nine have flat intrinsic spectra associated with thermal radio emission and the remaining 11 have steep intrinsic spectra, associated with synchrotron emission from supernova remnants. A supernova rate upper limit of 2.4 yr$^{-1}$ is determined for the inner 320 pc region of the galaxy at the 95\% confidence level, based on the lack of detection of new sources in observations spanning almost 17 years and a simple model for the evolution of supernova remnants. A supernova rate of $>0.14 (v/ 10^{4})$ yr$^{-1}$ is implied from estimates of supernova remnant source counts, sizes and expansion rates, where $v$ is the radial expansion velocity of the supernova remnant in km s$^{-1}$. A star formation rate of $3.4 (v/10^{4}) < SFR(M\geq5M_{\Sun})<59$ M$_{\Sun}$ yr$^{-1}$ has been estimated directly from the supernova rate limits and is of the same order of magnitude as rates determined from integrated FIR and radio luminosities.
\end{abstract}


\section{Introduction}
\label{sec:introduction}

Starburst galaxies are those galaxies containing regions (usually associated with the nucleus) currently undergoing high rates of star formation, much higher than seen in galaxies like our own and such that a significant fraction of the available gas for star formation will be exhausted within a single dynamical timescale \citep{leh96}.  A starburst is triggered by the rapid accumulation of gas in a small volume, causing large numbers of massive stars to form.  The accumulation of gas can be caused by interactions between galaxies, for example the merger of gas rich spiral galaxies can result in the accumulation of 60\% of the gas within the inner 100 pc of the merger product \citep{bar96}.  Gas can be funnelled into the nuclear region of a galaxy by dynamical processes associated with bar instabilities \citep{com00}.

Once formed, the starburst hosts large numbers of massive stars that ionise their environment and evolve quickly to form supernovae \citep{eng98}.  The supernova remnants can drive activity in the nuclear region and can produce strong winds out of the disk of the galaxy \citep{doa93}.

Starbursts are therefore the result of processes acting on a wide range of scales and themselves drive energetic activity on a wide range of scales.  By direct observation of the supernova remnants in starburst regions at radio wavelengths (to mitigate against severe extinction at optical wavelengths and provide very high angular resolution) it is possible to investigate the individual remnants and use them to reconstruct the supernova and star formation history of the starburst.  Furthermore, it is possible to use radio observations to investigate the ionised gaseous environment of the starburst.  Such observations provide a link between large-scale dynamical effects in the galaxies, activity in the star forming region itself, and the energetic phenomena in turn driven by the starburst. 

NGC 253 is a prominent nearly edge-on starburst galaxy in the Southern Hemisphere ($\delta\sim-25^{\circ}$), its proximity making it an ideal target for high spatial resolution observations. Previous studies of NGC 253 have assumed a distance of 2.5 Mpc \citep{tur85}, based on the estimate of \citet{dev78}. Both this and the revised estimate of 2.58$\pm$0.7 Mpc \citep{puc88} were derived from the Tully-Fisher relation. More recently \citet{kar03} determined the magnitude of the tip of the red giant branch from Hubble Space Telescope/WFPC2 images and arrived at a more reliable distance estimate of 3.94$\pm$0.5 Mpc. In this paper this new distance estimate is adopted.

In many respects NGC 253 is a twin of it's Northern Hemisphere counterpart M82. Both are of similar size, are nearly edge-on, barred, and have similar infrared spectra and luminosities. \cite{eng98} confirmed the existence of a large-scale bar $\sim$7 kpc in diameter crossing a circumnuclear ring of similar size in a de-reddened K-band image of NGC 253. Using a combination of optical, infrared and millimetre observations, \citet{for00} concluded that a ring of young star clusters $\sim$80 pc in diameter defined the inner edge of a cold gas torus centred on the assumed nucleus. The torus is fed by gas channelled in by the large-scale bar and deposited at the Inner Lindblad Resonance \citep{arn95}.  Australia Telescope Compact Array (ATCA) images of para-NH$_{3}$ (1,1) and (2,2) and ortho-NH$_{3}$ (3,3) and (6,6) were used to probe the temperature of the nuclear regions and highlight some of the inner structure of the torus where the temperatures are greatest \citep{ott05}. The full extent of the torus is seen more clearly in CO and [\ion{C}{1}] emission line maps which trace cool gas and dust \citep{isr95}, these observations showing that the torus is clumpy and appears to have a diameter of $\sim900$ pc.

Associated with the nuclear region of the galaxy is gas producing strong optical emission lines such as H$\alpha$, [\ion{S}{2}], and [\ion{O}{3}] \citep{for00,car94}.  The environment is heavily ionised and the emission line region has a complex morphology \citep{for00}.  Also associated with the nuclear region, and having its origin in the inner starburst region, is a powerful wind that has formed a bipolar nuclear outflow cone perpendicular to the disk of the galaxy. The outflow has been investigated by \citet{hec90} and \citet{sch92} using H$\alpha$ spectral line measurements but is seen more clearly in X-ray observations by Chandra \citep{wea02, str02}.  \ion{H}{1} plumes are also seen bordering the H$\alpha$ and X-ray halo emission and are also believed to be associated with the inner starburst region and active star formation in the disk \citep{boo05}. It is thought that the wind is matter blown from the central starburst region by the ensemble of massive stars and supernova remnants within the inner 80 pc.

\begin{figure}
\plotone{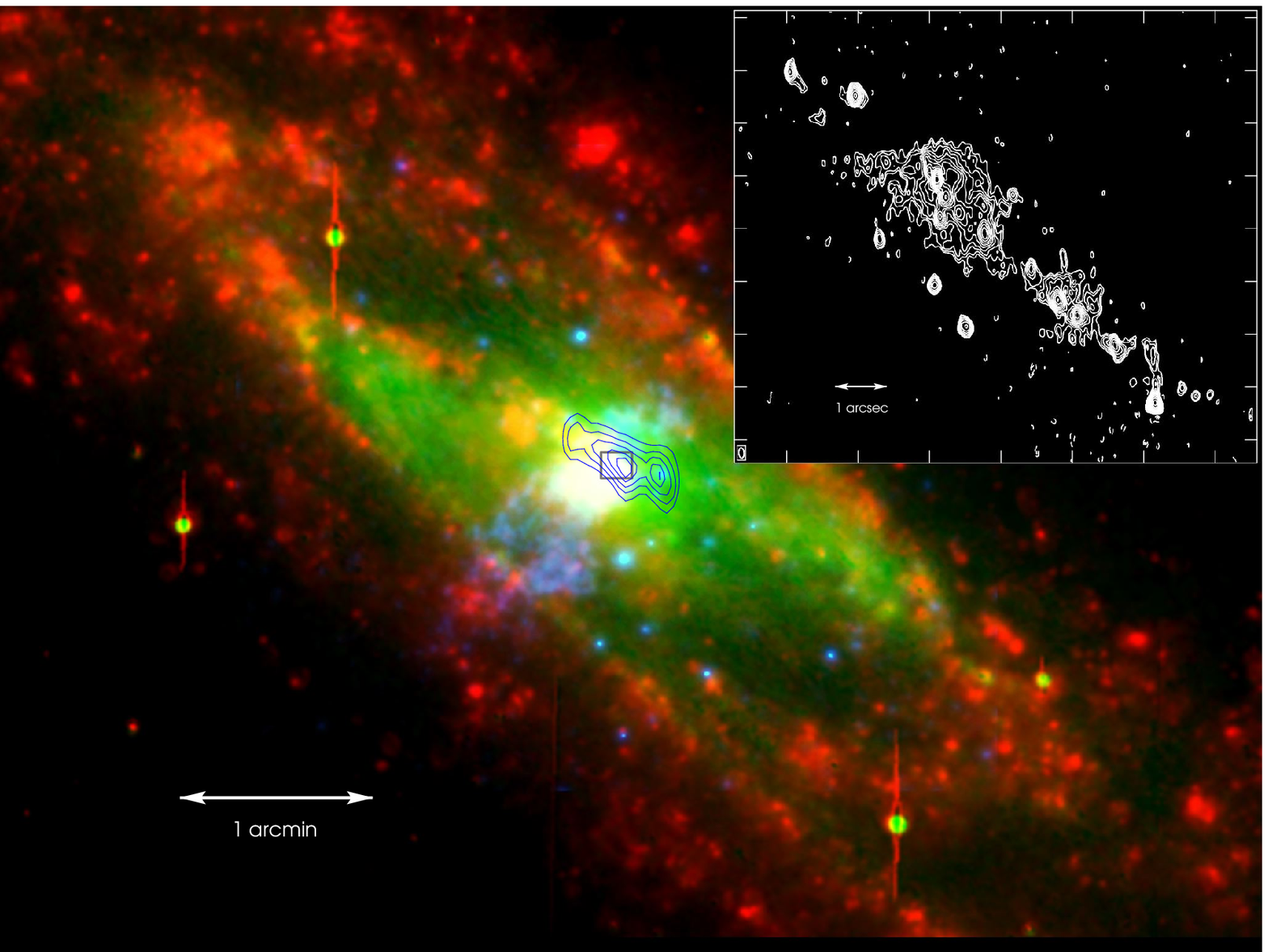}
\caption{A three-colour composite, multi-wavelength view of NGC 253. Red, green and blue indicate H$\alpha$ (courtesy of M. Dahlem), K$-$band 2MASS IR \citep{jar03}, and Chandra soft x-ray \citep{wea02} respectively. The Owens Valley Radio Interferometer CO emission line contour map of the inner 3 kpc is shown in blue with contours at 19\%, 38\%, 57\%, 76\% and 87\% of the peak (courtesy of M. Dahlem and F. Walter and first published by \citet{ott05}). Inset: VLA A configuration 2 cm image of central boxed region, contours are logarithmic intervals of $2^{1/2}$, beginning at 0.225 mJy beam$^{-1}$ (U\&A97).}
\label{fig:figmultiwav}            
\end{figure}

Figure \ref{fig:figmultiwav} presents images of the galaxy at various wavelengths, on different spatial scales, to illustrate of the relationships between the galaxy bar (IR), the molecular torus (CO), the inner starburst region (radio), and the wind eminating from that region (H$\alpha$ and x-ray).  The supernova remnants that are a primary subject of this paper are concentrated within the inner 80 pc of the galaxy but are also distributed widely outside the inner starburst region. 

At radio wavelengths, \citet[hereafter U\&A97]{ulv97} imaged NGC 253 with the VLA at frequencies between 1.5 GHz and 23 GHz over a period of eight years and identified 64 individual compact sources within 320 pc of the nucleus. The identification of sources at lower frequencies ($<10$ GHz) was hindered by confusion and the diffuse emission of the galaxy. Spectral indices, with an error less than 0.4, could therefore only be determined for 17 compact sources, mainly those isolated sources outside the nuclear region. In the absence of any new sources or any significant fading of existing sources over the course of their observations, U\&A97 derived a supernova rate of no more than 0.3 yr$^{-1}$. At 23 GHz the assumed nucleus (5.78-39.4) is the strongest source and is unresolved. Approximately half of the 17 detected compact sources with measured spectral indices were shown by U\&A97 to be associated with \ion{H}{2} regions. The remainder are most likely to be supernova remnants.

In the first attempt to resolve the inner starburst region of NGC 253 at low radio frequencies, \citet{tin04} used the Australian Long Baseline Array (LBA) in the first multi-baseline VLBI observation of NGC 253. At 1.4 GHz the observation approximately matched the angular resolution of the VLA at 23 GHz, thus overcoming confusion problems. Two sources were detected, neither of which corresponded to the assumed nucleus. \citet{tin04} showed that the radio emission from compact sources in NGC 253 is partially absorbed by ionised gas with free-free optical depths ranging between $\tau_{0}=2.5$ and $\tau_{0}>8$ ($\tau_{0}$ denotes the free-free optical depth at 1 GHz).  Further evidence for free-free absorption in NGC 253 comes from measurements of the nuclear starburst region between 300 MHz and 1.5 GHz by \citet{car96}, showing a dramatic spectral flattening at low frequencies. The modelling of radio-recombination line emission from the nuclear region of NGC 253 suggests that the lines originate from a gas with an electron density of between $\sim7\times10^{3}$ and $\sim1.7\times10^{4}$ cm$^{-3}$ and with a temperature of $7.5\times10^{3}$ K, distributed in a spherical structure of uniform density and a diameter of 3$-$4 pc \citep{moh02}. \citet{tin04} found that these results were consistent with a high degree ($\tau_{0} \sim 36$) of free-free absorption towards the assumed nucleus, 5.79$-$39.0.

Similar results for the supernova remnants in M82 have been obtained by \citet{ped99} and \citet{mcd02}.  In M82 30 supernova remnants have been detected from a population of 46 compact radio sources \citep{mcd02} and several resolved \citep{ped99, mux05}.  \citet{mcd02} found a free-free optical depth toward supernova remnants of $\tau_{0}\sim3$.

Motivated by the strong indicators of free-free absorption in NGC 253, we have embarked on a program of VLBI observations of this galaxy and other prominant Southern Hemisphere starburst galaxies. This paper reports results from a new observation of the nuclear region of NGC 253 at 2.3 GHz. The resulting data have better resolution and sensitivity than the 1.4 GHz observations  of \citet{tin04} and we have used them to investigate in more detail the free-free absorbed spectra of supernova remnants and \ion{H}{2} regions in the NGC 253 starburst (\S~\ref{sec:ffmodel}). In particular we are interested in probing the structure of the ionised environment of the supernova remnants and the relationship between the ionised medium and the other components of the galaxy (\S~\ref{sec:spectra}), further constraining the supernova rate in NGC 253 using previously published images and our new VLBI images (\S~\ref{sec:snrate}) and re-evaluating estimates of the star formation rate based on our results (\S~\ref{sec:sfrate}).

A subsidiary focus of this project is to achieve high sensitivity, high fidelity, high resolution, and computationally efficient imaging over wide fields of view (at least compared to traditional VLBI imaging techniques).  As such, the associated exploration of the relevant techniques is a small first step toward the much larger task of imaging with the next generation of large radio telescopes, for example the Square Kilometre Array (SKA: \citet{hal05, car04}).  The SKA will use baselines ranging up to approximately 3000 km and cover fields of view greater than one square degree instantaneously at frequencies of around 1 GHz.  The computational load required for this type of imaging task is vast and will drive the development of novel imaging algorithms, using techniques such as w-projection\footnote{EVLA Memo 67:  \url{http://www.nrao.edu/evla/geninfo/memoseries/evlamemo67.pdf}} and the use of super-computing facilities.  Investigating the performance of these techniques now, under the most challenging current observing conditions, is therefore a useful activity.

%
%
%
\section{Observations, data reduction, and results}

\subsection{Observations and data reduction}
\begin{figure}
\plotone{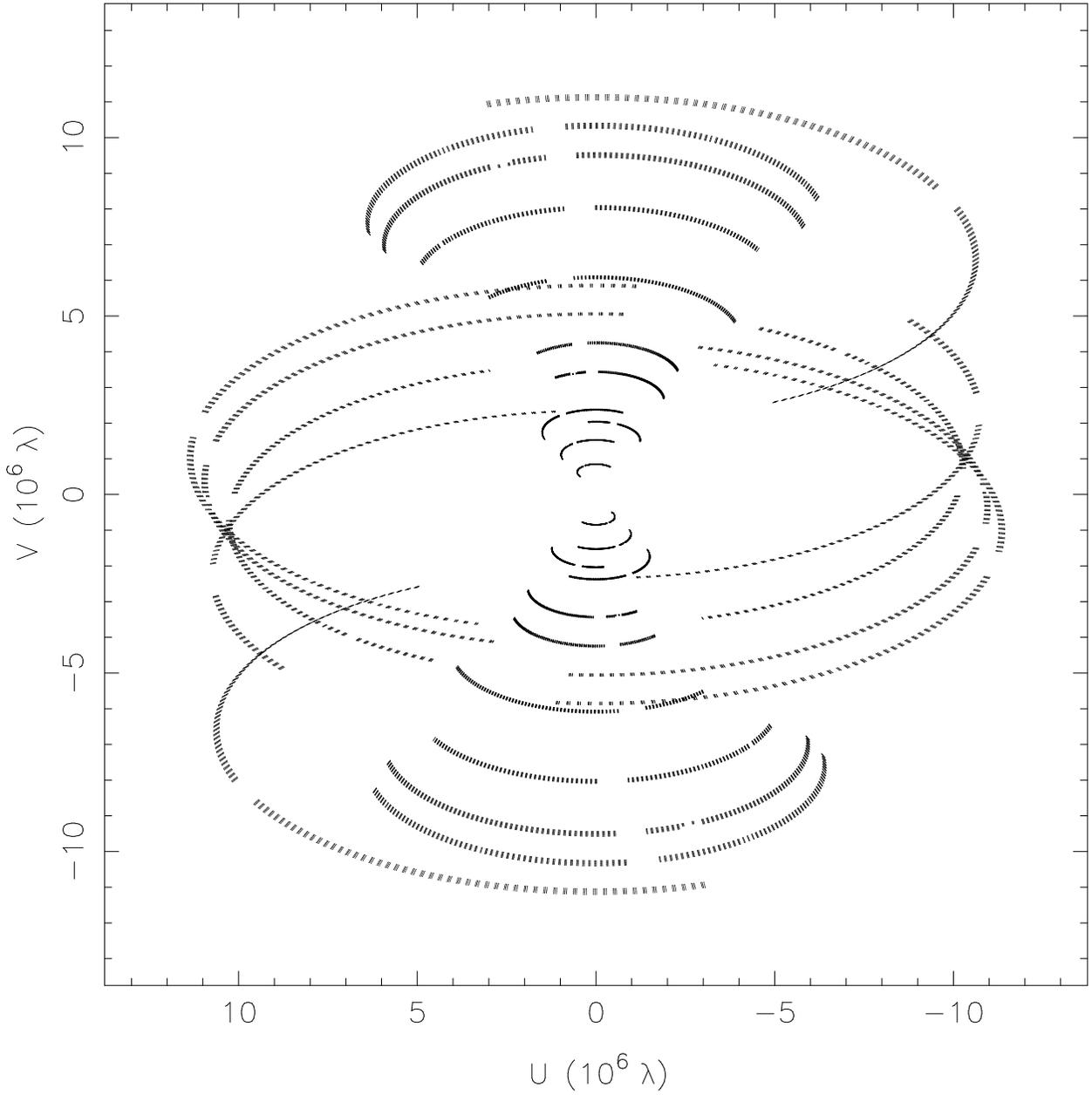}
\caption{Spatial frequency ($u,v$) coverage for NGC 253 at a frequency of 2.3 GHz.}
\label{fig:figuvcov}            
\end{figure}

A VLBI observation of NGC 253 was made on 16/17 April, 2004 (19:00 - 07:00 UT) using a number of Australian radio telescopes: the 70 m NASA Deep Space Network (DSN) antenna at Tidbinbilla; the 64 m antenna of the Australia Telescope National Facility (ATNF) near Parkes; 5 $\times$ 22 m antennas of the ATNF Australia Telescope Compact Array (ATCA) near Narrabri used as a phased array; the ATNF Mopra 22 m antenna near Coonabarabran; the University of Tasmania's 26 m antenna near Hobart; and the University of Tasmania's 30 m antenna near Ceduna.  The observation utilised the S2 recording system \citep{can97} to record 2 $\times$ 16 MHz bands (digitally filtered 2-bit samples) in the frequency ranges: 2252 - 2268 MHz and 2268 - 2284 MHz.  Both bands were upper side band and right circular polarisation.  The spatial frequency ($u,v$) coverage for the observation is shown in Figure \ref{fig:figuvcov}.

During the observation, three minute scans of NGC 253 ($\alpha = 00$:47:33.178; $\delta = -25$:17:17.060 [J2000]) were scheduled, alternating with three minute scans of a nearby phase reference calibration source, PKS J0038$-$2459 ($\alpha = 00$:38:14.735493; $\delta = -24$:59:02.23519 [J2000]).

The recorded data were correlated using the ATNF Long Baseline Array (LBA) processor at ATNF headquarters in Sydney \citep{wil92}. The data were correlated using an integration time of 2 seconds and with 32 frequency channels across each 16 MHz band (channel widths of 0.5 MHz).

The correlated data were imported into the AIPS\footnote{The Astronomical Image Processing System (AIPS) was developed and is maintained by the National Radio Astronomy Observatory, which is operated by Associated Universities, Inc., under co-operative agreement with the National Science Foundation} package for initial processing. The data for the phase reference source were fringe-fit (AIPS task FRING) using a one minute solution interval, finding independent solutions for each of the two 16 MHz bands.  The delay and phase solutions for the phase reference source were examined and averaged over each three minute calibrator scan, following editing of bad solutions, before being applied to both the phase reference source and NGC 253.  Further flagging of the data was undertaken via application of a flag file that reflected the times during which each of the antennas were expected to be slewing, or time ranges that contained known bad data.  Finally, data from the first 30 seconds of each scan from baselines involving the ATCA or Parkes were flagged, to eliminate known corruption of the data at the start of each scan at these two telescopes.

During correlation, nominal (constant) system temperatures (in Jy) for each antenna were applied to the correlation coefficients in order to roughly calibrate the visibility amplitudes (mainly to ensure roughly correct weights during fringe-fitting).  Following fringe-fitting, the nominal calibration was refined by collecting and applying the antenna system temperatures (in K) measured during the observation, along with the most recently measured gain (in Jy/K) for each antenna.  Further refinements to the amplitude calibration were derived from observations of a strong and compact radio source (PKS B1921$-$293) made immediately before the NGC 253 observations. PKS B1921$-$293 was observed using the same array as NGC 253, at the same frequencies and bandwidths, and simultaneously the PKS B1921$-$293 data were recorded at the ATCA. PKS B1921$-$293 is unresolved at this frequency and on these VLBI baselines, therefore the flux measured at the ATCA could be used to check and refine the amplitude calibration for the VLBI data.  The PKS B1921$-$293 data were also used to derive a bandpass calibration in the MIRIAD package \citep{mir95}, using task MFCAL, which was applied to the data for NGC 253 and the phase reference source.  The edge channels of each band were edited from the dataset (3 channels from the lower edge and 2 channels from the upper edge of each 32 channel band).

\begin{figure}
\plotone{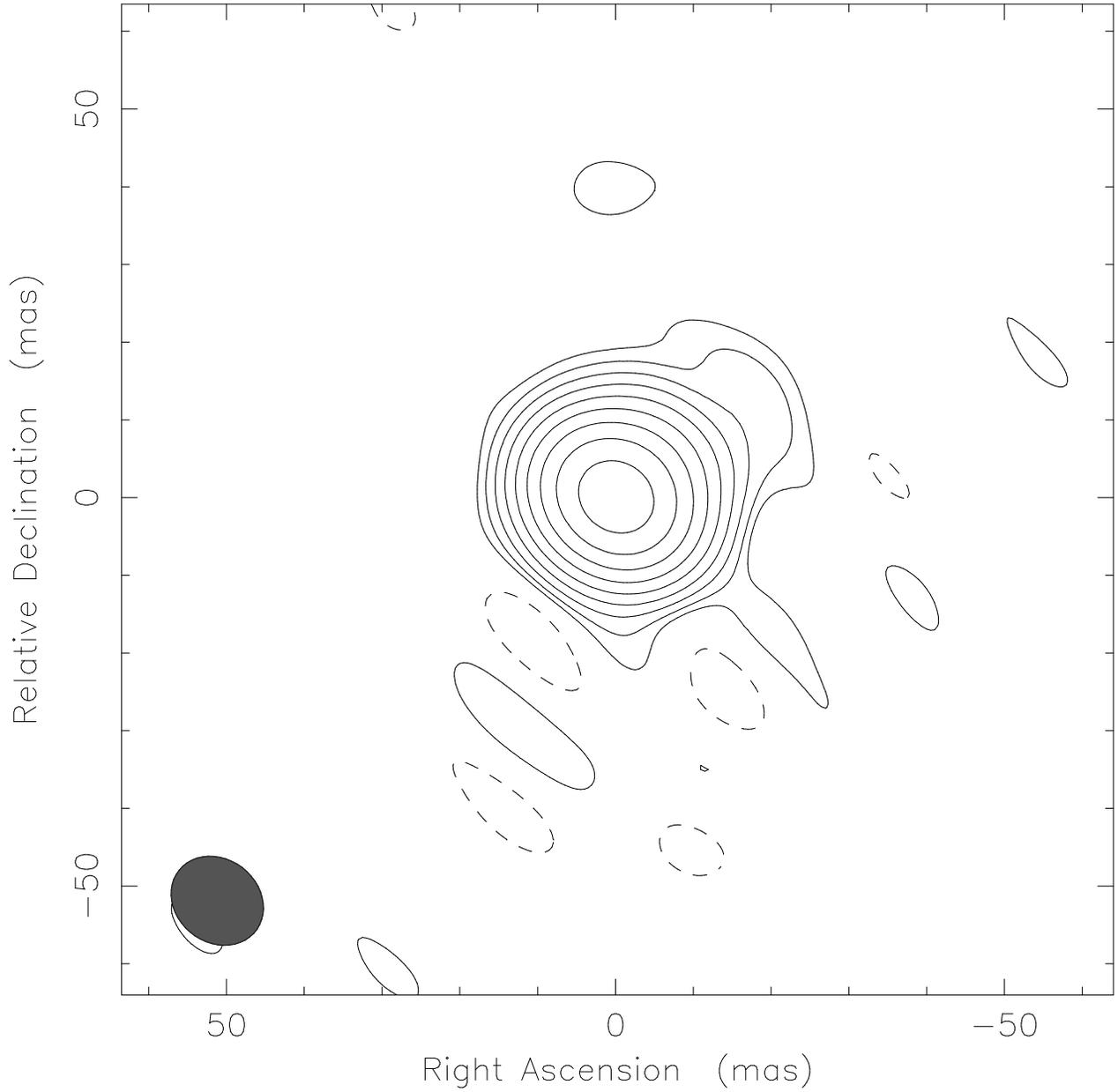}
\caption{Australian LBA image of the source used for phase calibration, PKS J0038$-$2459 at a frequency of 2.3 GHz. The image has an RMS noise level of approximately 0.25 mJy/beam. The peak is 361 mJy/beam and contours are at -0.25\%, 0.25\%, 0.5\%, 1\%, 2\%, 4\%, 8\%, 16\%, 32\% and 64\% of the peak. The beam size is approximately 13 $\times$ 11 mas at a position angle of $51^{\circ}$}
\label{fig:figj0038}            
\end{figure}

In DIFMAP \citep{she94} the data for the phase reference calibrator were vector averaged over a 30 second period and then imaged using standard imaging techniques (de-convolution and self-calibration of both phase and amplitude).  The resulting image (Figure \ref{fig:figj0038}) shows that PKS J0038$-$2459 is highly compact, with no significant structure on these baselines at this frequency. During self-calibration, amplitude corrections of less than 10\% were noted and applied to the NGC 253 data in a second iteration of the amplitude calibration in AIPS.  For the Tidbinbilla antenna, no system temperatures were measured, so a constant value as a function of time was initially applied. To refine the Tidbinbilla calibration a gain-elevation curve was derived in MIRIAD using the task MFCAL. This curve was applied to the data set to complete the amplitude calibration procedure.

\begin{figure}
\epsscale{0.7}
\centerline{\includegraphics[angle=270,width=\textwidth]{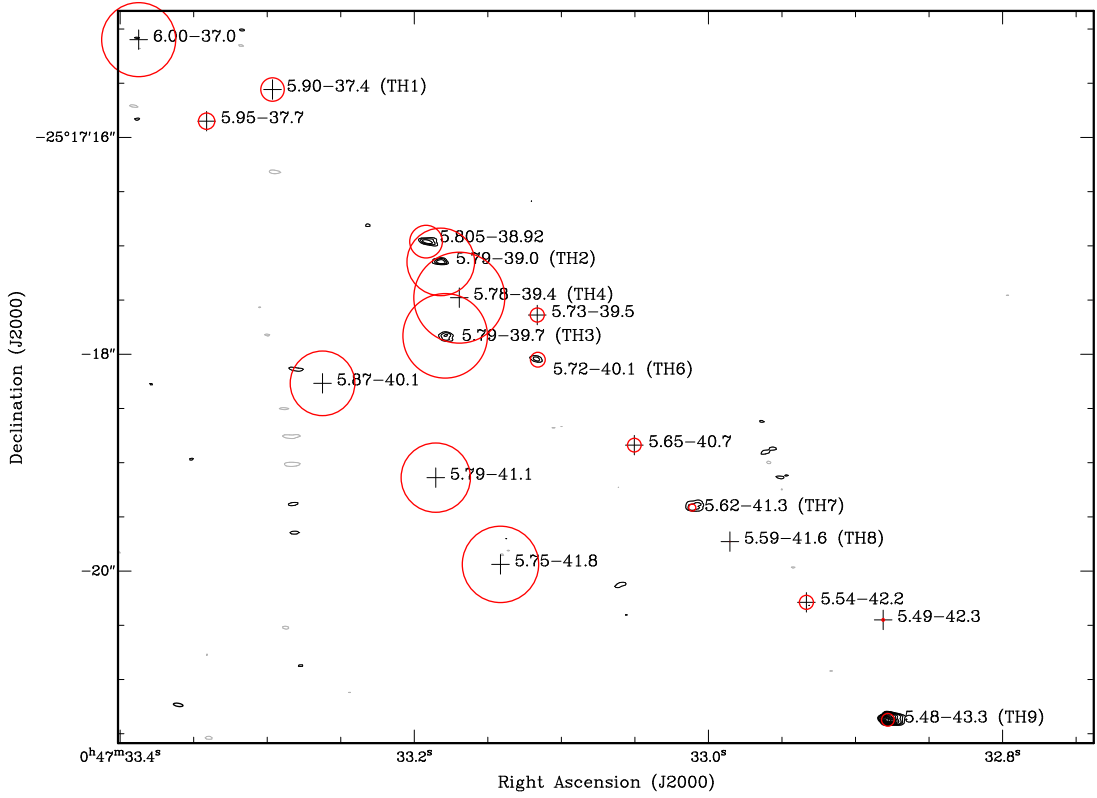}}
\caption{Australian LBA image of NGC 253 at a frequency of 2.3 GHz. The image has an RMS noise level of approximately 0.24 mJy/beam. The peak is 8.56 mJy/beam and contours are at -10\%, 10\%, 15\%, 20\%, 30\%, 40\%, 50\%, 60\%, 70\%, 80\%, 90\% and 100\% of the peak. The beam size is approximately 90 $\times$ 30 mas at a position angle of $81^{\circ}$. The overlayed circles give an indication of the degree of absorption in the vicinity of the source with the diameter of the circle being directly proportional to $\tau_{0}$ (or the lower limit on $\tau_{0}$). All sources are labelled using U\&A97 notation, \citet{tur85} sources are labelled within parentheses and non-detections are marked with a cross. }
\label{fig:figngc253}            
\end{figure}

Once amplitude calibrated, with the delay and phase solutions from fringe-fitting applied, and the bandpass calibration applied, the data for both phase reference source and NGC 253 were exported as FITS files.  Both data sets were examined in DIFMAP and some editing was undertaken. Imaging of the NGC 253 data took place in AIPS++\footnote{AIPS++ (Astronomical Information Processing System) was originally developed by an international consortium of observatories (ASTRON, ATNF, JBO, NCSA, NRAO) as a toolkit for the analysis/reduction of radio astronomical data. Development now continues based on project/instrument needs.}, using the w-projection algorithm. W-projection allows wide fields to be imaged, accounting for the w-term in such a way that imaging artefacts are at far lower levels than for other more traditional VLBI wide-field imaging techniques such as faceting \citep{per94}. Following a number of trial imaging runs, the parameters chosen for imaging were as follows: cell size = 11 mas; weighting = briggs and weighting robustness = 2. The chosen weighting scheme minimised noise at the expense of resolution but provided a beam shape that enabled sources to be identified more readily with the available spatial frequency coverage. Figure \ref{fig:figngc253} shows the image of NGC 253 using the data set described above but excluding the data from the Hobart and Ceduna antennas (the longest baselines).

\begin{figure}
\plotone{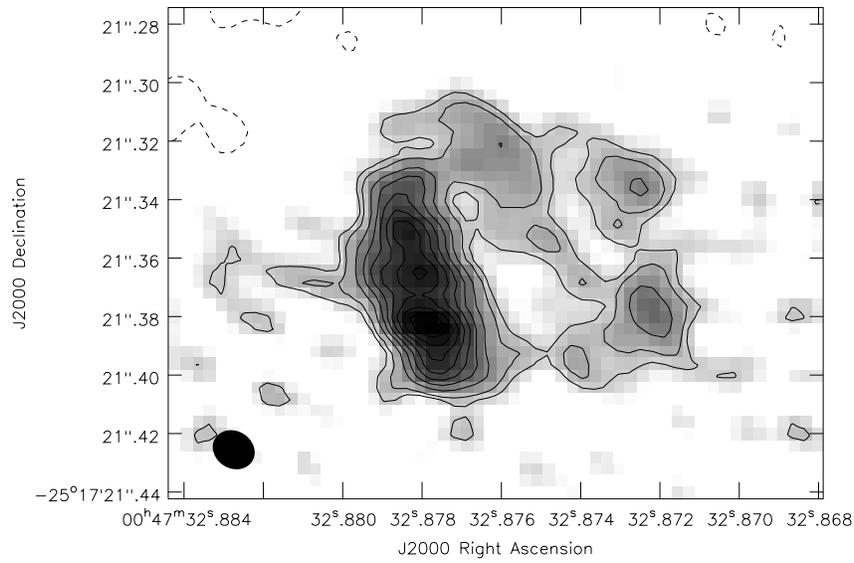}
\caption{Australian LBA image of the compact source 5.48$-$43.3 in NGC 253 at a frequency of 2.3 GHz. The image has an RMS noise level of approximately 0.16 mJy/beam. The peak is 2.42mJy/beam and contours are at -15\%, 15\%, 20\%, 30\%, 40\%, 50\%, 60\%, 70\%, 80\%, 90\% and 100\% of the peak. The beam size is approximately 15 $\times$ 13 mas at a position angle of $58^{\circ}$.}
\label{fig:figsnr}            
\end{figure}

In Figure \ref{fig:figngc253}, a number of compact radio sources have clearly been detected.  The one sigma RMS noise in Figure \ref{fig:figngc253} is 0.24 mJy/beam, a factor of 4 greater than the theoretically predicted thermal noise value\footnote{Estimated with the ATNF VLBI sensitivity calculator: \url{http://www.atnf.csiro.au/vlbi/calculator/}} for this array, bandwidth and observation time.  As there were no sufficiently bright sources to perform self-calibration, the higher noise level may be attributed to residual phase errors on time-scales less than the 3 minute calibrator-target duty cycle. Amplitude errors and limited spatial frequency coverage are also likely to have been contributing factors. We take a 5 sigma detection limit in this image of 1.2 mJy/beam. Figure \ref{fig:figsnr} is an image of 5.48$-$43.3 (bottom right corner in Figure \ref{fig:figngc253}) made from the full VLBI data set (including the Hobart and Ceduna data) and with similar imaging parameters used to produce \ref{fig:figngc253}, except that a smaller cell size of 4 mas was chosen.

\subsection{Identification of Sources}
\label{sec:sourceid}
A list of the sources detected above the 5 sigma detection threshold in the new VLBI data is given in Table \ref{tab:tabsources}. Flux density errors of $\pm10$\% are listed due to uncertainties in the absolute flux density scale for Southern Hemisphere VLBI \citep{rey94}.  The total flux density for each source was determined by summing its CLEAN model components in the image plane and an estimate of the source size, for all but 5.48$-$43.3, was obtained by fitting a Gaussian source to the CLEAN model components. As can be seen from the list, the brightest of the detected sources (5.48$-$43.3) is well resolved into a shell at this resolution. The shell diameter of 5.48$-$43.3 was measured by taking a cross-cut through the centre and measuring the distance between shell maxima. 

A comparison of the source positions with previous high resolution imaging of NGC 253 shows that each of the detected sources can be readily identified. Table \ref{tab:tabsources} lists identifications with 1.3 cm and 2 cm sources found by U\&A97 with VLA observations. Where available, identifications with 2 cm sources found by \citet{tur85}, also with the VLA, are specified with the ``TH$n$" notation originally adopted by U\&A97. All other compact radio sources in the NGC 253 starburst detected by U\&A97 at 23 GHz lie below our 1.2 mJy/beam detection limit at 2.3 GHz. \citet{tin04} sources (A) and (B), detected at 1.4 GHz with the LBA, are also identified with the U\&A97 sources 5.48$-$43.3 and 5.62$-$41.3 respectively but are not listed in the table.

We find a position difference of 71$\pm$53 mas between our J2000 source positions and the precessed B1950 positions of the 23 GHz sources from U\&A97. These differences reduce to 51$\pm$24 mas if the weakest source, 5.72$-$40.1, is excluded from the sample. We also find an offset of 94$\pm$13 mas between the 1.4 GHz LBA source positions of \citet{tin04} and our 2.3 GHz positions. A contribution to the difference between our derived positions and the positions listed by U\&A97 comes from the uncertainty in the phase calibrator positions used at the VLA during the NGC 253 observations of U\&A97. For example, the calibrator 0008$-$264, used by U\&A97 for their 1.3 cm, 2.0 cm, 3.6 cm, and 6.0 cm observations of 1987, 1989 and 1991, had an uncertainty of 150 mas at the time of its use as a calibrator. The calibrator used for our VLBI observations, J0038$-$2459, is used as part of the ICRF and has a quoted uncertainty of $\sim0.7$ mas \citep{bea02}. Our observations further benefited from a smaller beam size and an improved spatial frequency coverage compared to the 1.4 GHz LBA observation. We estimate our astrometric errors, based on half the minor axis of the beam, to be $\pm7$ mas for 5.48$-$43.3 and $\pm15$ mas for the remaining sources. 

\section{Discussion}

\subsection{Radio spectra and free-free absorption modelling of the compact sources}
\label{sec:ffmodel}
U\&A97 measured the spectral indices of 23 compact radio sources in NGC 253 using VLA observations between 5 GHz and 23 GHz (17 with spectral index errors of less than 0.4). Assuming that the spectral indices apply down to 2.3 GHz and that the sources are not resolved out by the higher resolution available with the LBA, 18 of these sources should be visible above the 5 $\sigma$ detection limit of our 2.3 GHz LBA observations. These sources are listed in Table \ref{tab:tabflux} with flux density measurements from VLA observations at 5 GHz, 8.3 GHz, 15 GHz and 23 GHz (U\&A97), together with LBA observations of two sources (5.48$-$43.3 and 5.62$-$41.3) at 1.4 GHz and flux density upper limits of 1.8 mJy at 1.4 GHz for the remaining sources \citep{tin04}.

Comparing this list with the sources in Table \ref{tab:tabsources} reveals that only 4 sources (5.48$-$43.3, 5.62$-$41.3, 5.72$-$40.1 and 5.79$-$39.0) were detected with the LBA at 2.3 GHz. The 2.3 GHz flux densities for all 18 sources are listed in Table \ref{tab:tabflux} with upper limits of 1.2 mJy for non-detections. A further two sources, 5.805$-$38.92 and 5.97$-$39.7, were detected with the LBA at 2.3 GHz and also by U\&A97 using the 15 GHz VLA A and 23 GHz VLA (A+B) configurations, and are also listed in Table \ref{tab:tabflux}. However, due to mismatches in resolution these source fluxes are only considered as approximate measurements by U\&A97. As a check to ensure that our measured fluxes at 2.3 GHz were not an underestimate as a result of higher resolution, total flux density measurements were compared between the 6 telescope LBA (Parkes, Narrabri, Mopra, Tidbinbilla, Hobart and Ceduna) observations and the 4 telescope array (excluding the longer baselines associated with Ceduna and Hobart) observations and were found to be the same within the measurement errors. Furthermore, U\&A97 found no significant variability in individual radio sources in NGC 253 over a period of 8 years, so it can be assumed that these measured fluxes, at multiple wavelengths, over multiple epochs, can be used to construct the spectra. 

\citet{tin04} argued that free-free absorption was the cause for the sharp downturn observed in the spectra of the compact radio sources at 1.4 GHz. At 2.3 GHz the effect of free-free absorption is not as great as at 1.4 GHz, so the detection of two sources at 1.4 GHz and a further four sources at 2.3 GHz appears to support this argument. The non-detection of the remaining 14 sources thus suggests that the effect of free-free absorption is still significant at 2.3 GHz. To test this quantitatively, an analysis similar to that undertaken by \citet{mcd02} for M82 was performed.

\citet{mcd02} constructed the spectra of 20 compact radio sources in M82 from existing and new observations between 1.42 GHz and 15 GHz using the VLA and MERLIN. The spectra were investigated using three models as described by the equations

\begin{equation}
S(\nu)=S_{0}\nu^{\alpha},
\end{equation}
\begin{equation}
S(\nu)=S_{0}\nu^{\alpha}e^{-\tau(\nu)},
\end{equation}
\begin{equation}
S(\nu)=S_{0}\nu^{2}(1-e^{-\tau(\nu)}),
\end{equation}
where
\begin{equation}
\tau(\nu)=\tau_{0}\nu^{-2.1}.
\end{equation}

Equation 1 represents a simple power-law spectrum, equation 2 a free-free absorbed power-law spectrum, and equation 3 a self-absorbed bremsstrahlung (free-free absorbed) spectrum. In these expressions $\alpha$ is the optically thin intrinsic spectral index, $\tau_{0}$ is the free-free optical depth at 1 GHz and $S_{0}$ is the intrinsic flux density of the source at 1 GHz. The combination of high frequency VLA data and 1.4 GHz LBA data, together with the detection of six sources at 2.3 GHz, and the more stringent upper limit for non-detections at 2.3 GHz, allow us to use these models to constrain the free-free parameters of the 20 compact radio sources listed in Table \ref{tab:tabflux}.

The spectrum of each source was tested against each of the three models described above using a reduced$-\chi^{2}$ criterion to determine the best fit. For all sources, the free-free absorbed power-law model produced a significantly better fit than the self-absorbed and power-law models, for example giving a mean reduced$-\chi^{2}$ fit of 0.8, 23, and 40 for models 2, 3 and 1 respectively. Upper limits of 1.8 mJy and 1.2 mJy were used where there were non-detections from the LBA observations at 1.4 GHz and 2.3 GHz, respectively. 

\begin{figure}
\epsscale{0.8}
\plotone{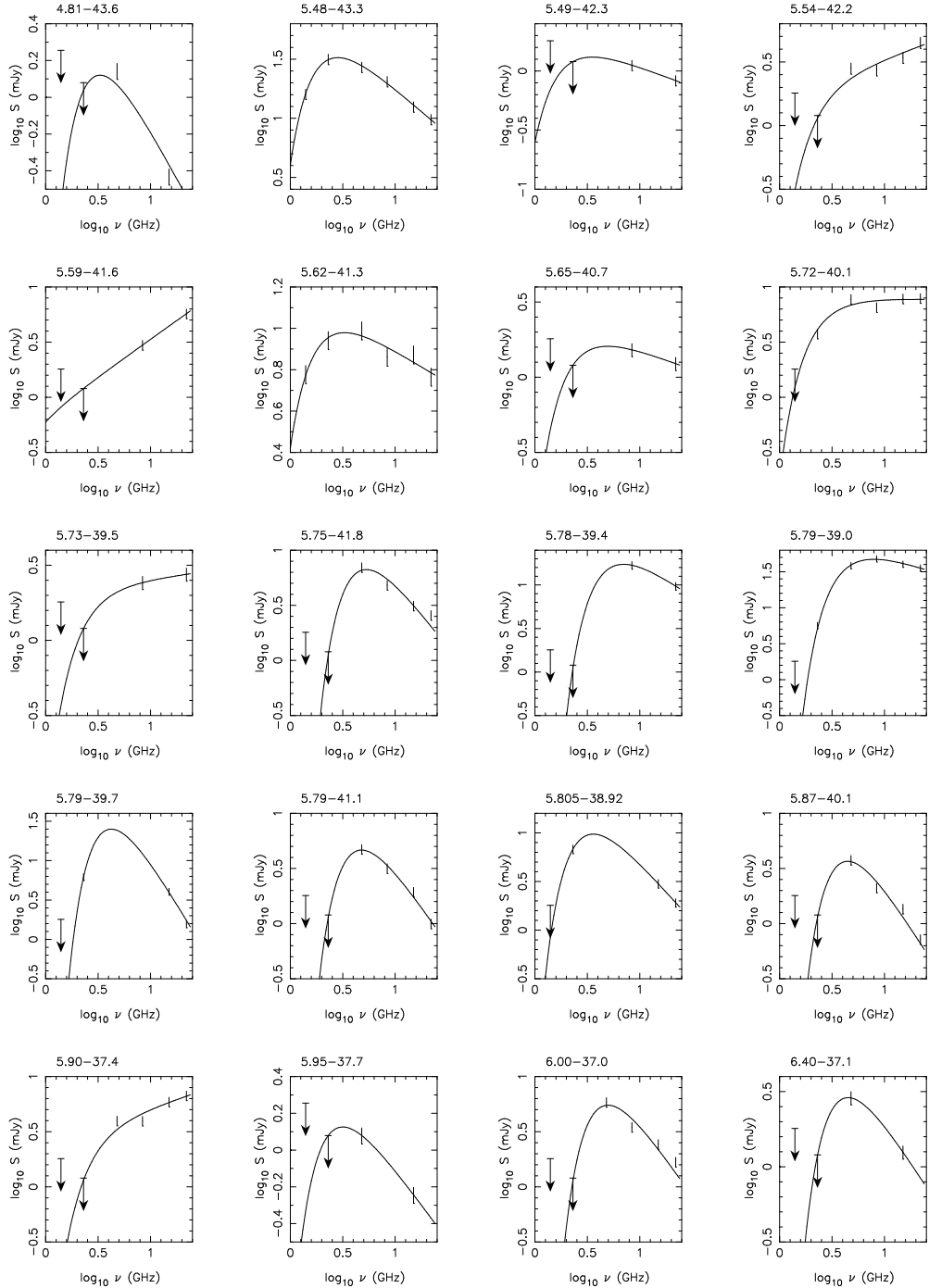}
\caption{Measured flux densities (points with error bars) and free-free absorption models (solid line) for 20 U\&A97 sources. The error bars are $\pm$10\% of the measured flux densities. 3$\sigma$ and 5$\sigma$ flux density upper limits are shown for sources not detected at 1.4 GHz (1.8 mJy) and 2.3 GHz (1.2 mJy) respectively.}
\label{fig:figff}            
\end{figure}

The resulting free parameters $S_{0}$, $\tau_{0}$ and $\alpha$ from the model fits are listed in Table \ref{tab:tabff}. The free-free absorbed power law model for each source is shown against the corresponding measured spectrum in Figure \ref{fig:figff}. The data shown in Table \ref{tab:tabff} and Figure \ref{fig:figff} confirm the free-free absorption interpretation of \citet{tin04}. In particular, the new 2.3 GHz data for 5.48$-$43.3 and 5.62$-$41.3 complete spectra that are very well fit by free-free absorbed power laws and are close to those adopted by \citet{tin04}. Furthermore, the results for the six sources listed in Table 2 of \citet{tin04} are consistent with the more stringently constrained free-free absorbed power law models listed in Table \ref{tab:tabff}.

Using similar criteria to that used by U\&A97, of the 20 sources listed in Table \ref{tab:tabff}, 9 have flat intrinsic power law spectra ($\alpha > -0.4$) indicative of \ion{H}{2} regions dominated by thermal radio emission (T), the 11 remaining sources have steep intrinsic spectra (S), as normally associated with supernova remnants \citep{mcd02}. The approximately equal number of thermal versus non-thermal sources is consistent with the findings of U\&A97 in NGC 253, however a slightly larger proportion of non-thermal sources is present in M82 \citep{mcd02}.

\subsection{Comparison with multi-wavelength datasets}
\label{sec:spectra}

When the modelled free-free opacity is illustrated against the source location (Figure \ref{fig:figngc253}) it is clear that it does not vary smoothly across the central region of the galaxy, but is rather clumpy. Since free-free absorption is associated with ionised gas, it is worthwhile investigating tracers of ionised gas at other wavelengths in an attempt to understand the distribution of gas in these regions.

HST images of NGC 253 at optical wavelengths provide the closest match in resolution to those obtained with the LBA and VLA. Of particular interest is the H$\alpha$ emission line as it traces diffuse ionised gas and is prominent in pre and post-main sequence stars. The \ion{S}{3} line has been used together with the H$\alpha$ line to highlight regions with photo-ionization from \ion{H}{2} regions and OB stars \citep{for00}. HST H$\alpha$ and [\ion{S}{3}] emission line images were made from archival HST data and the images shifted so that the source co-ordinates were aligned with those listed by \citet{for00}. The two images were then correlated against the free-free opacities modelled at the compact source locations. As many sources only had lower limits for the free-free optical depth, simple statistical techniques to test the relationship between measured intensity and the free-free opacity would not fully utilise all of the data available. Instead, 'survival analysis' methods were employed to take into consideration the lower limits when performing linear regression tests on the data.  The survival analysis was performed on the data using tasks in the ASURV package \citep{lav92}. The resulting correlation tests indicated that no trend was apparent between the H$\alpha$ or [\ion{S}{3}] emission line intensity and the modelled free-free optical depth for any of the sources. The lack of correlation is most likely due to the high levels of extinction that are apparent at optical wavelengths and are characteristic of starburst galaxies \citep{eng98}. This extinction makes properties deduced from optical measurements unrepresentative \citep[e.g.][]{rie80, pux91}.

To overcome the problems associated with extinction it is necessary to observe at wavelengths, such as X-ray and radio, that can penetrate gas and dust. Radio observations in particular also need to be toward the high frequency end of the spectrum to avoid the effects of free-free absorption. Australia Telescope Compact Array (ATCA) images of the Ammonia \citep{ott05} and HCN emission lines \citep{ott04} and \emph{Chandra} hard x-ray images \citep{wea02} were obtained and compared against the modelled free-free optical depth at each of the source locations. Ammonia (NH$_{3}$) is a good temperature indicator whereas HCN is a good tracer for high density molecular gas ($>10^{4}$ cm$^{-3}$). The hard x-ray observations show regions of photo-ionization by massive young stars or SNR-driven shocks. All three images were analysed using a survival analysis against the modelled free-free optical depth with no clear trend evident. Here the lack of correlation is most likely due to the mismatches in resolution. In comparison to the 2.3 GHz data, the ATCA and \emph{Chandra} images were of substantially lower resolution and so were not able to resolve sufficient detail to provide a useful comparison with the VLBI data.

High resolution radio recombination line (RRL) observations with the VLA provide a reasonable match in resolution to the VLBI observations. Furthermore, the high frequency H92$\alpha$ (8.3094 GHz) and H75$\alpha$ (15.2815 GHz) lines are not adversely affected by the effects of free-free absorption and can be used to estimate temperatures, densities, and kinematics \citep{lob04}. The hydrogen radio recombination line (RRL), which occurs in thermal gas, is insensitive to dust extinction and can be used to probe highly obscured star formation.  \citet{moh05} have imaged RRL emission in NGC 253 at arcsecond resolution and detected 8 sources. By modelling the emission, \citet{moh05} provide size and electron density estimates for these sources. 

When the RRL source associated with the core is overlaid with the core position in the 2.3 GHz LBA continuum map, 6 of the sources modelled for free-free absorption in Table \ref{tab:tabff} are found to be located near 5 of the RRL sources. By assuming that the sources are centrally located within a uniform density sphere of ionised gas of temperature 7500 K, the expected free-free opacities have been estimated, the results listed in Table \ref{tab:tabrrlff}.

Four of the VLA sources (5.49$-$42.3, 5.59$-$41.6, 5.72$-$40.1 and 5.78$-$39.4) have free-free opacities that are within the limits imposed by the RRL models (labelled as 'low' and 'high' in Table \ref{tab:tabrrlff}). The first three sources are identified as thermal regions. However 5.78$-$39.4 is non-thermal ($\alpha\sim-0.8$). The close proximity of 5.78$-$39.4 to the assumed core 5.79$-$39.0, where dense ionised gas is expected, together with the observed RRL emission, suggests that it may be a supernova remnant embedded in or behind a \ion{H}{2} region. 

The two VLA sources associated with the 5.795$-$39.05 RRL source, the assumed core (5.79$-$39.0) and a nearby \ion{H}{2} region (5.805$-$38.92) have free-free optical depths that fall well below the limits imposed by the RRL models. This suggests that the sources may not be centrally located within the gas distribution. If the sources are in front of or to the side of the high density region then they may exhibit the lower free-free optical depth that is inferred.

Many of the other VLA sources exhibit a significant degree of free-free absorption but do not have any associated RRL emission. Most of these sources are non-thermal and so may not necessarily have a RRL component. However, there are five thermal sources that do not have RRL sources associated with them. Four of these are weak and may fall below the sensitivity limit of the RRL observation. The remaining source has a combination of thermal and non-thermal emission (5.62$-$41.3) and it is possible that the thermal component is not strong enough to be detected within the sensitivity limits of the RRL observation. Interestingly, 5.62$-$41.3 does lie within a beam width of the RRL emission associated with 5.59$-$41.6 and were it associated with that emission it would have a free-free optical depth that is consistent with that expected from the source parameters modelled by \citet{moh05} from the RRL emission.

\subsection{Comments on individual compact radio sources}

\subsubsection{Resolved SNR 5.48$-$43.3}

Source 5.48$-$43.3 dominates observations at 1.4 GHz and 2.3 GHz. In our high resolution image (Figure \ref{fig:figsnr}) 5.48$-$43.3 appears to be resolved into a shell of approximately 90 mas in diameter corresponding to an extent of approximately 1.7 pc. If we assume an average expansion velocity of $v=10,000$ km s$^{-1}$ relative to the shell centre, similar to that measured for the supernova remnant 43.31+592 in M82 \citep{ped99}, this implies an age of $\sim 80 (10^{4}/v)$ years for this remnant. The eastern side of the remnant appears 2 to 3 times brighter than the western side. This may be the result of interaction with denser interstellar material in that direction. The circular symmetry of the remnant may indicate that the interaction has only recently begun or that the density of the ISM is only marginally greater.

With a total flux of 32 mJy at 2.3 GHz, 5.48$-$43.3 is approximately half the size and 25 times brighter than Cas A would be at this frequency if it were at the same distance. Furthermore, with the assumed expansion velocity, 5.48$-$43.3 is one quarter the estimated age of Cas A. 

\subsubsection{Assumed Core 5.79$-$39.0}

The measured size of 5.79$-$39.0 in the 2.3 GHz LBA image is approximately 80$\times$60 mas, consistent with the lower limit of 30 mas measured with the Parkes-Tidbinbilla Interferometer \citep{sad95} and the upper limit of $\sim100$ mas measured by U\&A97 with 8 GHz VLA observations. 5.79$-$39.0 is not detected in the full resolution LBA image at 2.3 GHz as would be expected given its size and flux density.

5.79$-$39.0 is the strongest source in 5$-$23 GHz VLA images but is not visible in 1.4 GHz LBA observations \citep{tin04}. In our 2.3 GHz LBA image the source is detected but is weaker than expected based on its high frequency spectral index, suggesting that there is significant absorption at lower frequencies. Our source modelling (\S\ref{sec:ffmodel}) shows that the source is best fit with a free-free absorbed power law model with optical depth $\tau_{0}\sim17$. Evidence of free-free absorption in the nuclear region has also been found by \citet{rey83}, \citet{car96} and \citet{tin04}.

\subsubsection{Resolved SNRs 5.62$-$41.3 and 5.79$-$39.7}

The source 5.62$-$41.3 lies in a region that has the highest HCN/CO ratio and thus the highest average density ($\sim10^{5}$ cm$^{-3}$) of any region studied by \citet{pag95}. This is one of only two sources detected at 1.4 GHz with the LBA \citep{tin04}. U\&A97 suggest that the source is a mix of supernova remnants and \ion{H}{2} regions. In our 2.3 GHz LBA observations the source is extended ($130\times85$ mas) in the low resolution image (Figure \ref{fig:figngc253}) and is not detected in the full resolution image. It does not appear to be greatly affected by free-free absorption and so may lie in front of the dense gas detected by \citet{pag95}. Its modelled spectral index of $-0.31$ is at the low end of that measured by U\&A97.

Source 5.79$-$39.7 is the smallest supernova remnant in our 2.3 GHz LBA image but is not detected in the higher resolution image. It also appears to be in a region with significant levels of ionised gas as it is heavily affected by free-free absorption.

\subsubsection{The \ion{H}{2} regions 5.72$-$40.1 and 5.805$-$38.92}
Source 5.72$-$40.1 is a flat spectrum source which has been observed to vary in its flux significantly at 5 GHz and 8.3 GHz from VLA observations (U\&A97). Our measured size of $60\times50$ mas is significantly smaller than the size of $230\times120$ mas measured at 23 GHz and $200\times90$ mas measured at 15 GHz using the VLA (U\&A97). It also has a position offset that is not as consistent as the other sources when compared against VLA positions. These two facts suggest that it may have structure that varies significantly with frequency. This source is associated with RRL emission that is consistent with a free-free absorbed \ion{H}{2} region.

Source 5.805$-$38.92 is extended in size ($130\times50$ mas) and not detected in higher resolution LBA images. It is associated with RRL emission but exhibits less free-free absorption than expected given the electron densities and source sizes predicted from RRL emission models.

\subsection{The supernova rate in NGC 253}
\label{sec:snrate}
With new observations made since those of U\&A97 and with an improved distance determination to NGC 253 \citep{kar03} it is worth reinvestigating the limits that may be placed on the supernova rate, discussed in detail by U\&A97.  The increased distance estimate to NGC 253 suggests that earlier spatial measures were underestimated by a factor of $\sim1.6$. Similarly, luminosity and energy estimates would have been underestimated by a factor of $\sim2.5$. The overall effect on the population of observed supernova remnants in NGC 253 is that they are not only older but also more luminous than previously estimated.

An upper limit to the supernova rate can be made by assuming that all of the FIR luminosity is reprocessed supernova energy, that is, all energy released by supernovae is eventually thermalized. Using this assumption \citet{ant88} determined an upper limit of 3 yr$^{-1}$ for the supernova rate. Adjusting the original FIR luminosity measured by \citet{tel80} for a distance of 3.94 Mpc gives a total FIR luminosity of $3.8\times10^{10} L_{\Sun}$ or $\sim1.4\times10^{44}$ ergs s$^{-1}$ for NGC 253. Dividing this by the canonical $10^{51}$ ergs of total energy output per supernova results in a maximum rate of $\sim4.5$ supernovae per year. This method has been shown to grossly overestimate supernova rates since FIR emission also results from stellar heating of dust and can be contaminated by the presence of an AGN \citep{con92}.

A lower limit on the supernova rate can be made based on the number of detected remnants, their size and an assumed expansion rate. \citet{ulv94} assumed that of order 100 compact sources were supernova remnants with upper size limits of 2$-$5 parsecs based on a distance of 2.5 Mpc to NGC 253. Furthermore, assuming an expansion at 5,000 km s$^{-1}$ (rate of expansion of the FWHM of a Gaussian model), \citet{ulv94} arrived at a supernova rate of 0.1$-$0.25 yr$^{-1}$. As only $\sim50\%$ of the compact radio sources in NGC 253 are associated with supernova remnants (U\&A97) and the revised source sizes are now of order 3$-$8 parsecs, a supernova rate of $>0.14-0.36$ is expected given an assumed expansion rate of 10,000 km s$^{-1}$. Since the mean expansion rate of the supernova remnants in NGC 253 is unclear, we can rewrite the limit as $\nu_{SN}>0.14 (v/10^{4})$ yr$^{-1}$, where $v$ is the shell radial expansion velocity in km s$^{-1}$. This is of the same order of magnitude as the supernova rate determined for M82, using similar methods, $\sim0.07$ yr$^{-1}$ \citep{ped03}.

Having observed NGC 253 at a second epoch, \citet{ulv91} found that no new sources stronger than 3 mJy at 5 GHz were detected after a period of 18 months. They statistically modelled the effect of such non-detections on the supernova rate. Their model assumed a hypothetical population of supernovae that peaked at 5 GHz 100 days after their optical maxima, had a 5 GHz peak flux that was with equal probability either 5, 10, 15, or  20 times that of Cas. A and decayed as the $-$0.7 power of time. \citet{ulv91} determined that approximately $2/3$ of supernovae that occur during the 18 month period between epochs should be detectable. Further assuming that the occurrence of supernovae in NGC 253 are Poisson distributed, \citet{ulv91} determined with 95\% confidence that the upper limit of the supernova rate was 3.0 yr$^{-1}$. The same model was used to determine an upper limit of $\sim1.4$ yr$^{-1}$ at the end of a third epoch \citep{ulv94} and $\sim0.3$ yr$^{-1}$ at the end of the fourth epoch (U\&A97).

Since the fourth epoch two further high resolution radio observations of NGC 253 have been made. The first was four years later by \citet{moh05} at 5.0 GHz and the second was the observation made 4.8 years later again at 2.3 GHz as described in this paper. Although the \citet{moh05} image can be directly compared to the historical data at the same frequency, the observations at 2.3 GHz require extra consideration of the nature of supernova remnants at lower frequencies, the differing sensitivity limits,  and the effects of free-free absorption. Regardless of these differences, what is clear is that no new sources are evident in either the \citet{moh05} images or our 2.3 GHz image.

To place limits on the supernova rate based on these observations, we have developed a new model, based in principle on that of \citet{ulv91}. At 5 GHz the hypothetical supernova for this model is identical to the one described in \citet{ulv91}, however at 2.3 GHz it is assumed that the supernova remnant will peak 200 days after the optical peak, albeit at the same flux density as at 5 GHz - this approximately follows the SN 1980k light curve determined by \citet{wei86}.

A Monte-Carlo simulation was used to drive the model by randomly choosing a uniformly distributed peak luminosity between 5 and 20 times that of Cas A., which was then scaled to flux density for a given distance to NGC 253 and also adjusted for free-free absorption using Eqn 2. A Poisson distributed random number was then used to define how many supernovae, given a specified supernova rate, would occur between two epochs and at which uniformly distributed times these supernovae would occur. Given the peak flux density and timing of the supernova, a test was made to determine if each supernova remnant could be detected when observed at the subsequent epoch given the sensitivity limit for the observation. By executing the Monte-Carlo simulation over 10,000 iterations, the proportion of supernova remnants detected at the end of each epoch was determined. The simulation was seeded with a supernova rate of 0.1 yr$^{-1}$ across all epochs, the resulting output giving the confidence level for a non-detection at each epoch at that rate. Linear interpolation was then used to determine a new rate that would drive the simulation towards a 95\% confidence level and the simulation was repeated until the 95\% level was reached.

Table \ref{tab:tabsnrate} shows the results of three separate tests with the Monte-Carlo simulation. In each test, a run was made for each epoch to place a limit on the supernova rate at the time of that observation. Also, given that a supernova remnants can peak and fade beneath the sensitivity limit between two epochs, the proportion of supernova remnants ($\beta_{SN}$) that can actually be detected at the end of each epoch was also noted.

In test 1 the distance to NGC 253 was set to 2.5 Mpc and no free-free absorption was assumed in order to provide a direct comparison to the \citet{ulv91, ulv94, ulv97} results.  At epoch 2, the first test resulted in a supernova rate of $<2.4$ yr$^{-1}$ compared to $<3.0$ yr$^{-1}$ obtained by \citet{ulv91}. This difference can be attributed to the proportion detected ($\sim83$\%) being greater than initially anticipated with a simpler model ($\sim67$\%). The proportion detected at epoch 3 ($\sim68$\%) is similar to that predicted by \citet{ulv91} however the cumulative effect of a greater rate of detection at the second epoch results in a supernova rate of only $<1.0$ yr$^{-1}$ compared to $<1.4$ yr$^{-1}$ obtained by \citet{ulv94}. At the fourth epoch a lower proportion of detections are predicted ($\sim48$\%) as a result of an increased number of supernova remnants fading below the sensitivity limits over the longer period between observations, this results in a supernova rate of $<0.62$ yr$^{-1}$ which is almost twice the maximum predicted by U\&A97. Given further observations by \citet{moh05} and those presented in this paper it is possible to add two further epochs to the test which in the absence of further detections results in a further lowering of the maximum allowable supernova rate to $<0.44$ yr$^{-1}$ and $<0.26$ yr$^{-1}$ for epochs 5 and 6 respectively. The results of our simulations show that the model originally used by \citet{ulv91} does not accurately determine the supernova rate when the time between epochs exceeds approximately 2 years, under the assumptions used.

In test 2 the Monte-Carlo simulation was repeated for NGC 253 at the revised distance of 3.94 Mpc. The increased distance results in more supernova remnants falling below the sensitivity limits of the observations. For observations at 5 GHz this results in the possibility of only 8$-$22\% of all supernovae occurring between epochs being detected. The overall effect is to increase the upper limit of the supernova rate to 3.0 yr$^{-1}$ at the end of epoch 4 and to 0.6 yr$^{-1}$ at the end of epoch 6.

It has been shown in \S~\ref{sec:spectra} that the supernova remnants in NGC 253 appear to be situated in or behind a screen of ionised gas. In test 3 the effects of free-free absorption on the upper limit of the supernova rate are taken into consideration with the Monte-Carlo simulation. For this test a median value of $\tau_{0}=6$ was taken based on free-free modelling in \S~\ref{sec:spectra}. The effect of free-free absorption at 5 GHz is not greatly significant but is sufficient to drop the proportion of detections to between 5\% and 13\%. At 2.3 GHz the effect of free-free absorption is more significant with the proportion of supernova remnants that can be detected dropping from 70\% to 10\%. Overall, this results in upper limits on the supernova rate of 5.2 yr$^{-1}$ at epoch 4 and 2.4 yr$^{-1}$ at epoch 6.

The overall supernova rate upper limit of 2.4 yr$^{-1}$ is below the upper limit determined from FIR emission, however it is more than an order of magnitude greater than that determined from source counts and sizes alone. U\&A97 observed no significant source fading over a period of 8 years and on this basis they could further limit the supernova rate to no greater than 0.3$-$0.6 yr$^{-1}$. The apparent lack of significant variability on time-scales of many years has also been observed for a majority of compact radio sources in M82 by \citet{kro00}. It is possible that the dense environment around the nuclear regions of NGC 253 and M82 inhibit the expansion and fading of the remnants. If this is the case then the remnants may be older than they appear and this could reduce the supernova rate further. It is also possible that we are only seeing the tip of the iceberg, there may be a large background supernova rate but we only observe the rare bright events.

To improve the supernova rate estimates further it would be useful to refine the Monte-Carlo simulator to include a more realistic probability distribution for the supernova remnant peaks and make better use of the \citet{wei86} radio supernova light curves. A reduction in the upper limit of the supernova rate can be achieved by increasing the probability of detecting new supernovae as they occur. This may be achieved with frequent high sensitivity observations of NGC 253. Furthermore, through multi-epoch observations, it should be possible to measure the rate of expansion of the brighter more resolved remnants and so allow for age determination at a greater level of accuracy and subsequently improved estimates of the supernova rate. 

\subsection{The star formation rate in NGC 253}
\label{sec:sfrate}

The star-formation rate (SFR) of a star-forming galaxy is directly proportional to its radio luminosity $L_{\nu}$ at wavelength $\nu$ \citep{con92, haa00}:
\begin{equation}
\left( \frac{SFR(M\geq5M_{\Sun})}{M_{\Sun}\mathrm{yr}^{-1}} \right)=Q \left \{ \frac{ \frac{L_{\nu}}{\mathrm{W Hz^{-1}}}}{\left[ 5.3\times10^{21} \left( \frac{\nu}{\mathrm{GHz}}\right) ^{-0.8} + 5.5\times10^{20} \left( \frac{\nu}{\mathrm{GHz}} \right) ^ {-0.1} \right]} \right \}.
\end{equation}
\citet{con92} derives this relation purely from radio considerations by calculating the contribution of synchrotron radio emission from SN remnants and of thermal emission from \ion{H}{2} regions to the observed radio luminosity. The factor $Q$ accounts for the mass of all stars in the interval $0.1-100 M_{\Sun}$ and has a value of 8.8 if a Saltpeter IMF ($\gamma=2.5$) is assumed - this IMF is assumed for the remainder of this section. \citet{ott05} used a measure of the 24 GHz continuum emission in NGC 253 to derive a SFR which, when adjusted for a distance of 3.94 Mpc, gives a SFR of 10.4$\pm$1 M$_{\Sun}$ yr$^{-1}$.

A large proportion of the bolometric luminosity of a galaxy is absorbed by interstellar dust and re-emitted in thermal IR. As a result FIR emission is a good tracer for young stellar populations. By modelling the total energy emission of a massive star and assuming that the contribution of dust heating by old stars is negligable, \citet{con92} determined the following relation between the star-formation rate of a galaxy and the FIR luminosity $L_{IR}$:

\begin{equation}
\left( \frac{SFR(M\geq5M_{\Sun})}{M_{\Sun}\mathrm{yr}^{-1}} \right)=9.1\times10^{-11}L_{FIR}/L_{\Sun}.
\end{equation}
Using \citet{rad01} FIR measures for NGC 253, adjusted for a distance of 3.94 Mpc, gives a SFR of 1.8$-$2.8 M$_{\Sun}$ yr$^{-1}$ for the inner $\sim300$ pc nuclear region and 3.5$-$4.3 M$_{\Sun}$ yr$^{-1}$ for the entire galaxy.

Finally, the star formation rate can be determined from the supernova rate directly. For most galaxies the total radio non-thermal and thermal luminosities and the FIR/radio ratio is proportional to the average star formation rate for $M\geq5M_{\Sun}$ \citep{con92}. Furthermore, as all stars more massive than $8M_{\Sun}$ become radio supernova, the star formation rate can be determined using the following relationship: 

\begin{equation}
\left( \frac{SFR(M\geq5M_{\Sun})}{M_{\Sun}\mathrm{yr}^{-1}}\right) \sim 24.4 \times \left[ \frac{\nu_{SN}}{\mathrm{yr}^{-1}}\right].
\end{equation}
Using the supernova rate limits of $0.14 (v/10^{4})<\nu_{SN}<2.4$ yr$^{-1}$ (\S~\ref{sec:snrate}) the limits on the star formation rate are $3.4 (v/10^{4}) < SFR(M\geq5M_{\Sun})<59$ M$_{\Sun}$ yr$^{-1}$. This is of the same order of magnitude as the nuclear SFR determined from FIR emissions and in agreement with that determined from 24 GHz continuum emissions alone. Using equations 6 and 7, the supernova rate can be estimated from the FIR-based SFR estimate giving a value of 0.07$-$0.11 yr$^{-1}$. This figure is in line with measurements based on source sizes and suggests that the simple model used to determine the supernova rate from FIR emission in \S~\ref{sec:snrate} was grossly overestimating the supernova rate. It is encouraging that the three separate estimates of the SFR in NGC 253 all produce results that are only a factor of a few different. These estimates are similar to star-formation rates derived from the supernova rate and FIR emission in M82 which give values of $\sim1.8$ and $\sim2.0$ M$_{\Sun}$ yr$^{-1}$ respectively \citep{ped03}.

\section{Summary}

We have imaged NGC 253 at 2.3 GHz using the LBA to produce the highest resolution image of the nuclear starburst region of this galaxy to date. We find the following results:
\begin{itemize}
\item Six compact radio sources were detected and identified against higher frequency VLA observations (U\&A97). Two of these sources are also identified in a 1.4 GHz LBA image \citep{tin04}.
\item In the highest resolution image (13 mas beam), the supernova remnant 5.48$-$43.3 is resolved into a shell-like structure approximately 90 mas (1.7 pc) in diameter with the eastern limb substantially brighter than the west. Assuming an average radial expansion velocity of $v=10,000$ km s$^{-1}$, the remnant is estimated to be approximately $80 (10^{4}/v)$ years of age.
\item By combining flux density measures from 1.4 GHz LBA, 2.3 GHz LBA and high frequency VLA observations, with upper flux density limits for non-detections, the spectra for 20 compact radio sources were determined. The spectra of these sources were fit by a free-free absorbed power law model.
\item 11 of the 20 sources have steep intrinsic spectra normally associated with supernova remnants, the nine remaining sources have flat intrinsic power law spectra ($\alpha>-0.4$) indicative of \ion{H}{2} regions.
\item Based on the modelled free-free opacities of 20 sources, the morphology of the ionised medium in the central region of NGC 253 is complex and clumpy in nature.
\item Multi-wavelength comparisons of the free-free opacity against optical tracers of ionised gas, such as H$\alpha$ and [\ion{S}{3}] emission line images, show no significant correlation. The lack of correlation here is most likely due to the characteristically high levels of interstellar extinction associated with starbursts.
\item Multi-wavelength comparisons of the free-free opacity against non-optical tracers of ionised gas, such as x-ray and ammonia and HCN emission line images, also failed to show any significant correlation. The lack of correlation here is attributed to the  substantially lower resolution of these images compared to the LBA images.
\item A comparison with radio recombination line images show that four of the modelled sources have free-free optical depths expected by RRL models.
\item A supernova rate upper limit of 2.4 yr$^{-1}$ in the inner 320 pc region of NGC 253 was derived from the absence of any new sources, taking into consideration the improved distance measure to the galaxy, a median free-free opacity and the sensitivity limits of 6 observations over a period of 16.8 years. A supernova rate of $>0.14 (v/ 10^{4})$ yr$^{-1}$ has been estimated based on an estimate of supernova remnant source counts, their sizes and their expansion rates.
\item Both upper and lower limits on the supernova rate could be further constrained with more frequent, high sensitivity observations with the LBA or the VLBA.
\item A star formation rate of $3.4 (v/10^{4}) < SFR(M\geq5M_{\Sun})<59$ M$_{\Sun}$ yr$^{-1}$ has been estimated directly from supernova rate limits for the inner 320 pc region of the galaxy.
\end{itemize}
%
%





\section*{Acknowledgments}

We thank Michael Dahlem for the provision of the H$\alpha$ and CO data, Niruj R. Mohan for the RRL data, Juergen Ott for the HCN and Ammonia data, and Kim A. Weaver for the X-Ray data. E.L. acknowledges support from a Swinburne University of Technology Chancellor's Research Scholarship, a CSIRO Postgraduate Student Research Scholarship and ATNF co-supervision. The Australia Telescope is funded by the Australian Commonwealth Government for operation as a national facility managed by the CSIRO. This research has made use of the NASA/IPAC Extragalactic Database (NED), which is operated by the Jet Propulsion Laboratory, California Institute of Technology, under contract with the National Aeronautics and Space Administration. We thank an anonymous referee for helpful comments on the manuscript, especially with regards to our comparison of VLA and LBA source positions.

\clearpage

\clearpage

\begin{table}
\begin{center}
\begin{tabular}{llclllllll} \hline \hline
\multicolumn{2}{c}{Position\tablenotemark{a}}             & & \multicolumn{2}{c}{Identification} &      &      & \multicolumn{3}{c}{Source Size} \\
                      \cline{1-2}                                      \cline{4-5}                                         \cline{8-10}
$\alpha$(J2000) & $\delta$(J2000) & &                      &                          &  Peak\tablenotemark{d}      & $S_{2.3}$\tablenotemark{e} & Major & Minor & P.A. \\
00:47                  & -25:17                & & U\&A97\tablenotemark{b}            & T\&H85\tablenotemark{c}                  & (mJy/beam)      &  (mJy) & (mas) & (mas) & (deg) \\ \hline \hline
32.878                & 21.372               & & 5.48$-$43.3     & TH9        & 8.6$\pm0.9$      & 32$\pm3$      & $90\pm7$ & $90\pm7$ & 0 \\
33.011                & 19.414               & & 5.62$-$41.3     & TH7        & 1.7$\pm0.2$      & 8.8$\pm0.9$  & $130\pm15$ & $90\pm15$ & 98 \\
33.116                & 18.050               & & 5.72$-$40.1     & TH6        & 1.5$\pm0.2$      & 3.8$\pm0.4$  & $60\pm15$ & $50\pm15$ & 47 \\
33.179                & 17.830               & & 5.79$-$39.7     & TH3        & 1.8$\pm0.2$      & 6.2$\pm0.6$  & $50\pm15$ & $60\pm15$ & 128 \\
33.182                & 17.148               & & 5.79$-$39.0     & TH2        & 2.4$\pm0.2$      & 5.7$\pm0.6$  & $80\pm15$ & $60\pm15$ & 141 \\
33.192                & 16.961               & & 5.805$-$38.92 & $\cdots$ & 2.1$\pm0.2$      & 6.8$\pm0.7$  & $130\pm15$ & $50\pm15$ & 78 \\ \hline
\tablenotetext{a}{The measured 2.3 GHz LBA source positions.}
\tablenotetext{b}{Source identifications with 1.3 cm and 2 cm sources found by U\&A97.}
\tablenotetext{c}{Source identifications with 2 cm sources found by \citet{tur85} using the ``TH$n$" designation of U\&A97.}
\tablenotetext{d}{The measured peak flux density for the sources at 2.3 GHz.}
\tablenotetext{e}{The measured flux density for the sources at 2.3 GHz.}
\end{tabular}
\caption{Compact sources detected with 2.3 GHz VLBI.}
\label{tab:tabsources}
\end{center}
\end{table}

\begin{table}
\begin{center}
\begin{tabular}{lrrrrrr} \hline \hline
      & \multicolumn{2}{c}{LBA\tablenotemark{b} (mJy)} & \multicolumn{4}{c}{VLA\tablenotemark{c} (mJy)}                  \\
              \cline{2-3}                 \cline{4-7}
VLA ID\tablenotemark{a}      & ~~~~$S_{1.4}$  & ~~~~$S_{2.3}$  & ~~~~$S_{5}$    & ~~~~$S_{8.3}$  & ~~~~$S_{15}$  & ~~~~$S_{23}$  \\ \hline \hline 
4.81$-$43.6 &  $<$1.8    & $<$1.2     &  1.39       & $\cdots$ & 0.37       & $\cdots$ \\
5.48$-$43.3 &  16           & 32         &  27.10     & 20.49      & 12.51     &  9.79     \\
5.49$-$42.3 &  $<$1.8    & $<$1.2     &  $\cdots$ & 1.12       & $\cdots$ &  0.83     \\
5.54$-$42.2 &  $<$1.8    & $<$1.2     &  2.82       & 2.73        &  3.42      &  4.48     \\
5.59$-$41.6 &  $<$1.8    & $<$1.2     &  $\cdots$ & 2.97       & $\cdots$ &  5.72     \\
5.62$-$41.3 &  6             & 8.8           &  9.77       & 7.29        & 7.47       &  5.84     \\
5.65$-$40.7 &  $<$1.8    & $<$1.2     &  $\cdots$ & 1.52       & $\cdots$ &  1.23     \\
5.72$-$40.1 &  $<$1.8    & 3.8           &  7.74       & 6.53        & 7.82       &  7.88     \\
5.73$-$39.5 &  $<$1.8    & $<$1.2     &  $\cdots$ & 2.42       & $\cdots$ &  2.75     \\
5.75$-$41.8 &  $<$1.8    & $<$1.2     &  6.98       & 4.83        & 3.13       &  2.57     \\
5.78$-$39.4 &  $<$1.8    & $<$1.2     &  $\cdots$ & 16.86     & $\cdots$ &  9.69     \\
5.79$-$39.0 &  $<$1.8    & 5.7           &  38.64     & 47.95      &  40.32    &  35.79    \\
5.79$-$39.7 &  $<$1.8    & 6.2           &  $\cdots$ & $\cdots$ & $\sim$4.1 &  $\sim$1.6 \\
5.79$-$41.1 &  $<$1.8    & $<$1.2     &  4.73       & 3.17        &  1.94      &  1.00     \\
5.805$-$38.92 & $<$1.8 & 6.8           &  $\cdots$ & $\cdots$ & $\sim$3.0 &  $\sim$1.9 \\
5.87$-$40.1 &  $<$1.8    & $<$1.2     &  3.77       & 2.11        &  1.36      &  0.72     \\
5.90$-$37.4 &  $<$1.8    & $<$1.2     &  3.97       & 3.96        &  5.89      &  6.72     \\
5.95$-$37.7 &  $<$1.8    & $<$1.2     &  1.20       & $\cdots$  &  0.57      & $\cdots$ \\
6.00$-$37.0 &  $<$1.8    & $<$1.2     &  5.85       & 3.48         &  2.43      &  1.68     \\
6.40$-$37.1 &  $<$1.8    & $<$1.2     &  2.87       & $\cdots$  &  1.25      & $\cdots$ \\ \hline
\tablenotetext{a}{``VLA ID" is the identification from U\&A97.}
\tablenotetext{b}{$S_{1.4}$ is the measured flux density from the LBA \citep{tin04}. $S_{2.3}$ is the measured 2.3 GHz flux density from this work. Upper limits on the flux density at 1.4 GHz and 2.3 GHz are provided for sources that have not been detected.}
\tablenotetext{c}{$S_{5}$, $S_{8.3}$, $S_{15}$ and $S_{23}$ are the measured 5 GHz, 8.3 GHz, 15 GHz and 23 GHz flux densities from the VLA respectively (U\&A97).}
\end{tabular}
\caption{A summary of flux density measurements for all available sources.}
\label{tab:tabflux}
\end{center}
\end{table}

\begin{table}[h]
\begin{center}
\begin{tabular}{lrrrrl} \hline \hline
Source	&  $S_{0}$ (mJy) &   $\alpha$	&  $\tau_{0}$	& Type \\ \hline \hline
4.81$-$43.6 & 7.78		& -1.06		& $>$6.2     & S \\
5.48$-$43.3 & 104.7		& -0.77		& 3.3	   & S \\
5.49$-$42.3 & 2.34		& -0.33		& $>$2.2     & T    \\ 
5.54$-$42.2 & 1.66		& 0.30		& $>$3.6     & T    \\
5.59$-$41.6 & 0.70		& 0.68		& $>$0.16   & T    \\ 
5.62$-$41.3 & 16.0		& -0.31		& 1.8	   & T    \\
5.65$-$40.7 & 2.73		& -0.26		& $>$3.5     & T    \\ 
5.72$-$40.1 & 8.02		& -0.01		& 3.8	   & T    \\
5.73$-$39.5 & 2.05		& 0.10		& $>$3.6   & T    \\ 
5.75$-$41.8 & 93.2		& -1.23		& $>$20	& S \\
5.78$-$39.4 & 125.0		& -0.82		& $>$23	& S \\
5.79$-$39.0 & 155.9		& -0.47		& 17		& S \\
5.79$-$39.7 & 1694.0        & -2.21		& 22		& S \\
5.79$-$41.1 & 79.6		& -1.39		& $>$18	& S \\
5.805$-$38.92 & 78.9	& 1.20		& 8.4	& T    \\
5.87$-$40.1 & 71.7		& -1.51		& $>$17	& S \\
5.90$-$37.4 & 2.52		& 0.32		& $>$6.0   & T    \\
5.95$-$37.7 & 4.74		& -0.78		& $>$4.2   & S \\
6.00$-$37.0 & 95.8		& -1.37		& $>$19	& S \\
6.40$-$37.1 & 25.1		& -1.09		& $>$12	& S \\ \hline \hline
\end{tabular}
\caption{Parameters of the free-free absorption models for all compact sources, as discussed in the text. Thermal sources are shown with type T and non-thermal synchrotron sources are shown as type S.}
\label{tab:tabff}
\end{center}
\end{table}

\begin{table}[h]
\begin{center}
\begin{tabular}{llccccccc} \hline \hline
                        &                      &                   &  \multicolumn{2}{c}{$n_{e}$}                     & \multicolumn{2}{c}{$l$} & \multicolumn{2}{c}{$\tau_{0}$} \\
                                                                          \cline{4-5}                                                    \cline{6-7}                        \cline{8-9}
RRL                 & VLA              &                   & low                          & high                         & low   & high   & low            & high           \\
Source             & Source         & $\tau_{0}$ & (10$^{3}$ cm$^{-3}$) & (10$^{3}$ cm$^{-3}$)  & (pc)  & (pc)    &                  &                  \\ \hline \hline
5.49$-$42.3     & 5.49$-$42.3 & $>$2.2       & 4                             & 10                            & 2      & 3.5     & 7.8             & 85             \\
5.59$-$41.6     & 5.59$-$41.6 & $>$0.16     & 2                             & 20                            & 1      & 5        & 1.0             &  490          \\
5.72$-$40.1     & 5.72$-$40.1 & 3.8             & 2                             & 10                            & 1.4   & 2.7     & 1.4             &   66           \\ 
5.75$-$39.9     & $\cdots$       & $\cdots$    & 2                             & 10                            & 1.7   & 2.9     & 1.7             & 71             \\
5.753$-$38.95 & $\cdots$       & $\cdots$    & 2                             & 10                            & 1.8   & 5        & 4.9             &  44            \\ 
5.773$-$39.54 & 5.78$-$39.4 & $>$23        & 0.9                          & 6                              & 2.6   & 6.4     & 1.3             & 32             \\
5.777$-$38.98 & $\cdots$       & $\cdots$    & 8                             & 10                            & 2      & 2.5     & 31              & 61             \\ 
5.795$-$39.05 & 5.805$-$38.92 & 8.4         & 7                             & 10                            & 2.5   & 2.5     & 30              & 61             \\
5.795$-$39.05 & 5.79$-$39.0 & 17              & 7                             & 10                            & 2.5   & 2.5     & 30              & 61             \\ \hline \hline
\end{tabular}
\caption{Comparison of free-free opacity determined from source spectrum modelling and from radio recombination line modelling.}
\label{tab:tabrrlff}
\end{center}
\end{table}

\begin{table}
\begin{center}
\begin{tabular}{llllllllll} \hline \hline
           &               &                      &  & \multicolumn{2}{c}{Test 1\tablenotemark{a}} & \multicolumn{2}{c}{Test 2\tablenotemark{b}} & \multicolumn{2}{c}{Test 3\tablenotemark{c}} \\
                                                                          \cline{5-6}                                                    \cline{7-8}                        \cline{9-10}
Epoch & Time    & $\nu_{obs}$  & Sensitivity   & $\beta_{SN}$ & $\nu_{SN}$   & $\beta_{SN}$ & $\nu_{SN}$ & $\beta_{SN}$    & $\nu_{SN}$  \\ 
           & (yr)       & (GHz)            & (mJy)          &                   & (yr$^{-1}$) &  & (yr$^{-1}$) &  & (yr$^{-1}$) \\ \hline \hline
1         & $\cdots$& 5.0               & 3.0               & $\cdots$    & $\cdots$ & $\cdots$& $\cdots$& $\cdots$& $\cdots$ \\  
2         & 1.5        & 5.0                & 3.0              & 0.828         & $<2.4$     & 0.220         & $<9.1$   & 0.128          & $<16$       \\
3         & 2.5        & 5.0                & 3.0              & 0.678         & $<1.0$     & 0.132         & $<4.5$   & 0.077          & $<7.8$      \\
4         & 4.0        & 5.0                & 3.0              & 0.477         & $<0.62$   & 0.083         & $<3.0$   & 0.048          & $<5.2$      \\
5         & 4.0        & 5.0                & 3.0              & 0.477         & $<0.44$   & 0.083         & $<2.3$   & 0.048          & $<3.9$      \\
6         & 4.8        & 2.3                & 1.2              & 0.993         & $<0.26$   & 0.697         & $<0.64$ & 0.100          & $<2.4$      \\  \hline
\tablenotetext{a}{Monte Carlo test run with a distance of 2.5 Mpc.}
\tablenotetext{b}{Monte Carlo test run with a distance of 3.94 Mpc.}
\tablenotetext{c}{Monte Carlo test run with a distance of 3.94 Mpc and a free-free opacity of $\tau_{0}=6.0$.}
\end{tabular}
\caption{The supernova rate upper limit based on Monte Carlo simulations run over six observing epochs. The time between epochs, the observing frequency and sensitivity of the observation are listed. At the end of each epoch the proportion of supernova remnants detected ($\beta_{SN}$) in that epoch is listed together with an estimate of the supernova rate upper limit based on all observations prior to and including that epoch. }
\label{tab:tabsnrate}
\end{center}
\end{table}

\end{document}